\documentclass[aps,pra,print]{revtex4}
\usepackage{graphicx}
\usepackage{amsfonts}
\usepackage{amssymb}
\usepackage{amsmath}

\begin{document}

\title{Trapping of two-component matter-wave solitons by mismatched optical
lattices}
\author{Z. Shi$^1$, K.J.H. Law$^1$ P.G.\ Kevrekidis$^{1}$, and B.A. Malomed$%
^2$}
\affiliation{$^{1}$ Department of Mathematics and Statistics,University of Massachusetts,
Amherst MA 01003-4515, USA \\
$^2$ Department of Physical Electronics, School of Electrical Engineering,
Faculty of Engineering, Tel Aviv University, Tel Aviv 69978, Israel}

\begin{abstract}
We consider a one-dimensional model of a two-component Bose-Einstein
condensate in the presence of periodic external potentials of opposite
signs, acting on the two species. The interaction between the species is
attractive, while intra-species interactions may be attractive too [the
system of the bright-bright (BB) type], or of opposite signs in the two
components [the \textit{gap-bright} (GB) model]. We identify the existence
and stability domains for soliton complexes of the BB and GB types. The
evolution of unstable solitons leads to the
%spontaneous symmetry breaking (SSB) in different forms, and
establishment of oscillatory states. The increase of the strength of the
nonlinear attraction between the species results in \textit{symbiotic
stabilization} of the complexes, despite the fact that one component is
centered around a local maximum of the respective periodic potential.
\end{abstract}

\maketitle

%\begin{document}

\section{Introduction}

Optical lattices (OLs) offer a powerful and ubiquitous tool for the creation
and control of various patterns in Bose-Einstein condensates (BECs). OLs are
induced by the interference of counterpropagating coherent laser beams
illuminating the condensate and creating an effective periodic potential for
atoms \cite{Markus}. A well-established fact is that OLs support \textit{gap
solitons} in BECs with repulsive interactions between atoms. Effectively
one-dimensional (1D) gap solitons were predicted \cite{GS} and then created
experimentally in BECs filling a ``cigar-shaped" trap, which acted in the
combination with an OL induced in the axial direction. In self-attractive
media, periodic potentials allow one to capture a soliton at a prescribed
position, and also give rise to stable multi-soliton complexes. In fact, the
model of a medium with the cubic self-attractive nonlinearity and effective
periodic potential applies not only to BEC but also to optics, where it
predicts spatial solitons in a planar waveguide with a transverse modulation
of the local refractive index. Actually, the model was first put forward in
the latter context, and fundamental solitons, as well as their bound states,
were found in it \cite{Wang}. Later, the same model was introduced in the
context of the mean-field dynamics in BECs \cite{Konotop}.

Another topic of great interest to BECs is the study of binary condensates,
which are most typically generated as mixtures of two different hyperfine
atomic states, with opposite values of the $z$-projection of the atomic spin
\cite{binary}. The interactions between atoms belonging to the same \cite%
{Feshbach} and different \cite{inter-Feshbach} hyperfine species can be
controlled (including a possibility to reverse the sign of the interaction)
by means of the Feshbach resonance. In particular, one may consider a binary
condensate that features intra-species repulsive interactions combined with
attraction between the species. It was predicted that the latter setting may
give rise to \textit{symbiotic} soliton complexes \cite{symbiotic}, which
are held together by the attraction overcoming the intrinsic repulsion.

Specific types of BEC solitons have also been predicted in settings assuming
a binary condensate trapped in OLs. In particular, the interplay of the
repulsion between the two species, if combined with the dynamical effect
induced by the lattice (an effective negative mass of collective
excitations) may give rise to 1D and 2D \textit{symbiotic gap solitons} \cite%
{gap-symbio,gap-symbio2}, even in the case when the intra-species
interaction is switched off. The addition of the intra-species repulsion
helps to expand the stability region of the symbiotic gap solitons \cite%
{gap-symbio}, while the interplay of the OL effects, intra-species
attraction, and repulsion between the two species gives rise to other new
types of solitons complexes, such as \textit{semi-gap} ones, with one of the
components belonging to the semi-infinite gap in the OL-induced linear
spectrum, while the other component sits in a finite bandgap \cite%
{gap-symbio2}. The case of the attraction between two self-repulsive species
was recently considered in Ref. \cite{Warsaw}, where it was demonstrated
that the attraction leads to a counter-intuitive result, \textit{viz}.,
\emph{splitting} between gap solitons formed in each species. This effect is
explained by a\ negative effective mass of gap solitons, which is one of
their principal characteristic features \cite{GS}.

In fact, the OL potentials acting on two species in a binary BEC need not be
identical. For example, if the actual source of the potential is not an OL,
but rather a periodically nonuniform distribution of a magnetic field
(acting in direction $z$ and modulated along $x$), which couples to the
atomic spin (while the basic trap is optical), then the effective potential
will, obviously, have opposite signs for the atomic states with opposite
values of the $z$-component of the spin. An issue of straightforward
interest is then to consider two-component soliton complexes supported by
pairs of \textit{mismatched lattices}. In particular, effects of the
mismatch on two-component 1D and 2D gap solitons with \emph{linear} coupling
between the components (which corresponds to BEC loaded into two parallel
tunnel-coupled traps) were studied in Refs. \cite{Arik}. It was found that
the mismatch affects the \textit{spontaneous symmetry breaking} (SSB)
bifurcation that accounts for a transition from symmetric or antisymmetric
solitons to asymmetric ones: the symmetric/antisymmetric states are replaced
by their \textit{quasi}-symmetric/antisymmetric counterparts, but the
bifurcation still occurs. An exception is the limit case when the phase
shift between the lattices is $\pi $ (i.e., the two OL potentials are
mutually opposite) -- then, the bifurcation is replaced by a \textit{%
pseudo-bifurcation}, in which branches of quasi-symmetric/antisymmetric and
asymmetric solutions asymptotically approach each other, but never meet \cite%
{Arik}. Effects of the mismatch between two gratings on the SSB bifurcations
of optical gap solitons were also studied in the model of two linearly
coupled parallel fiber Bragg gratings, which is an optical counterpart of
the OL pair in the BEC model \cite{Yossi} (``light bullets" in an array of
Bragg gratings with alternating signs were considered in Ref. \cite{Canberra}%
).

The objective of the present work is to study 1D two-component solitons with
the ordinary nonlinear \emph{attraction} between the species, assuming that
they are subject to the action of periodic potentials \emph{with opposite
signs}. The model corresponds, in particular, to the above-mentioned mixture
of two atomic states with opposite values of the spin, that couples to a
periodically modulated distribution of the magnetic field. Issues of
straightforward interest are the existence and stability of two-component
solitons. The stability problem is quite nontrivial in this setting, as, on
the one hand, the soliton complex tends to be unstable because one of its
constituent components is placed around a local maximum of the respective
potential, but, on the other hand, the complex may be stabilized by the
attraction between the components. We consider two varieties of the model,
with either intrinsic attraction in both components [the respective soliton
complexes are categorized as \textit{bright-bright} (BB) ones], or opposite
signs of the intra-species interaction [which corresponds to \textit{%
gap-bright} (GB) complexes].

It is necessary to mention that various solutions (chiefly of the solitary
wave type) in
models of two-component BEC mixtures trapped in OLs, with attractive inter-
and intra-species interactions, were considered in several works \cite%
{misc,OK}. In most cases, the periodic potentials acting on both species
were essentially identical (in-phase) \cite{misc}, although an example with
the phase shift of $\pi /2$ between the two potentials was considered too
\cite{OK} (the latter work was chiefly dealing with soliton complexes of the
dark-bright type). However, the settings studied in the present work were
not considered before, to the best of our knowledge.

The rest of the paper is organized as follows. We formulate the model in
Sec. II, and summarize results for both types of the soliton complexes, of
the BB and GB types, in Sec. III. Stability regions in a relevant parameter
space are identified through the numerical computation of eigenvalues for
small perturbations. %; for
%the BB model, a simple analytical approximation is developed too.
The evolution of unstable solitons is explored by means of systematic direct
simulations. The paper is concluded by Sec. IV.

\section{The model}

The starting point is the system of coupled 1D Gross-Pitaevskii equations
for the mean-field wave functions of the two BEC species, $U_{1}$ and $U_{2}$
\cite{Pit}. In the usual scaled form, the system is%
\begin{equation}
i\frac{\partial U_{j}}{\partial t}=-\frac{1}{2}\frac{\partial ^{2}U_{j}}{%
\partial x^{2}}+\sum_{k=1}^{2}g_{jk}|U_{k}|^{2}U_{j}+V_{j}(x)U_{j},
\label{eqn1}
\end{equation}%
with $j,k=1,2$, where $g_{jk}\equiv 4a_{jk}\sqrt{m\Omega /\hbar }$ are
effective interaction coefficients ($a_{jk}$ are scattering lengths of the
respective interatomic interactions, $m$ the atomic mass, common for both
species, and $\Omega $ the transverse trapping frequency), and $V_{j}(x)$
are the normalized OL potentials, which are different for the two
components. In accordance with the scaling, the numbers of atoms in the two
species (their norms) are $N_{j}=\int_{-\infty }^{+\infty }\left\vert
U_{j}(x)\right\vert ^{2}dx$. As usual, we seek for stationary solutions to
Eqs. (\ref{eqn1}) with chemical potentials $\mu _{j}$ as $U_{j}(x,t)=\exp
(-i\mu _{j}t)u_{j}(x)$, where the real functions $u_{j}$ obey the equations
\begin{equation}
\mu _{j}u_{j}=-\frac{1}{2}\frac{d^{2}u_{j}}{dx^{2}}%
+\sum_{k=1}^{2}g_{jk}u_{k}^{2}u_{j}+V_{j}(x)u_{j}.  \label{eqn2}
\end{equation}

Localized solutions to Eqs. (\ref{eqn2}) were obtained through fixed-point
(Newton-Raphson) iterations, using initial guesses for both components in
the form of \textrm{sech} functions. As indicated in the introduction, the
situations that we focus on here are of the BB and GB types, i.e., ones with
self-attraction in both components ($g_{11,}g_{22}<0$), or opposite signs of
the self-interaction ($g_{11}>0$ and $g_{22}<0$), respectively.

Once the numerical solution to Eq. (\ref{eqn2}) was found, its stability was
investigated against infinitesimal perturbations $\left\{
a_{j},b_{j}\right\} $ with growth rate $\lambda \equiv \lambda _{\mathrm{real%
}}+i\lambda _{\mathrm{imag}}$, taking the perturbed solution as
\begin{equation}
U_{j}(x,t)=\exp (-i\mu_{j}t)\left[ u_{j}+a_{j}(x)\exp (\lambda
t)+b_{j}^{\star }(x)\exp (\lambda ^{\star }t)\right] ,  \label{eqn3}
\end{equation}%
and solving the eigenvalue problem generated by the substitution of this
expression in Eqs. (\ref{eqn1}) and linearization. Sets of unstable complex
eigenvalues of the resulting Hamiltonian eigenvalue problem arise in
quartets, i.e., if $\lambda $ is an eigenvalue, then so are $-\lambda $, $%
\lambda ^{\star }$ and $-\lambda ^{\star }$, hence the existence of any $%
\lambda _{\mathrm{real}}\neq 0$ implies instability (real eigenvalues come
in pairs, $\pm \lambda $, which also gives rise to instability). The
eigenvalue problem was solved by using a finite-difference discretization
[in the same way as it was used to solve the stationary version of Eq. (\ref%
{eqn1})]. The latter method reduces the problem to finding matrix
eigenvalues, which can be done by means of standard numerical linear algebra
packages.

If a configuration is found to be unstable, an outcome of the instability
development was concluded from direct simulations. To that effect, we used
an integrator implemented on the basis of the fourth-order Runge-Kutta
method, with initial conditions taken as per exact numerically found
stationary states, with the addition of weak perturbations seeding the
instability growth.

\section{Results: two-component solitons and their stability}

\subsection{Bright-bright solitons}

\begin{figure}[tbp]
\begin{center}
\includegraphics[width=7.3cm,height=6cm]{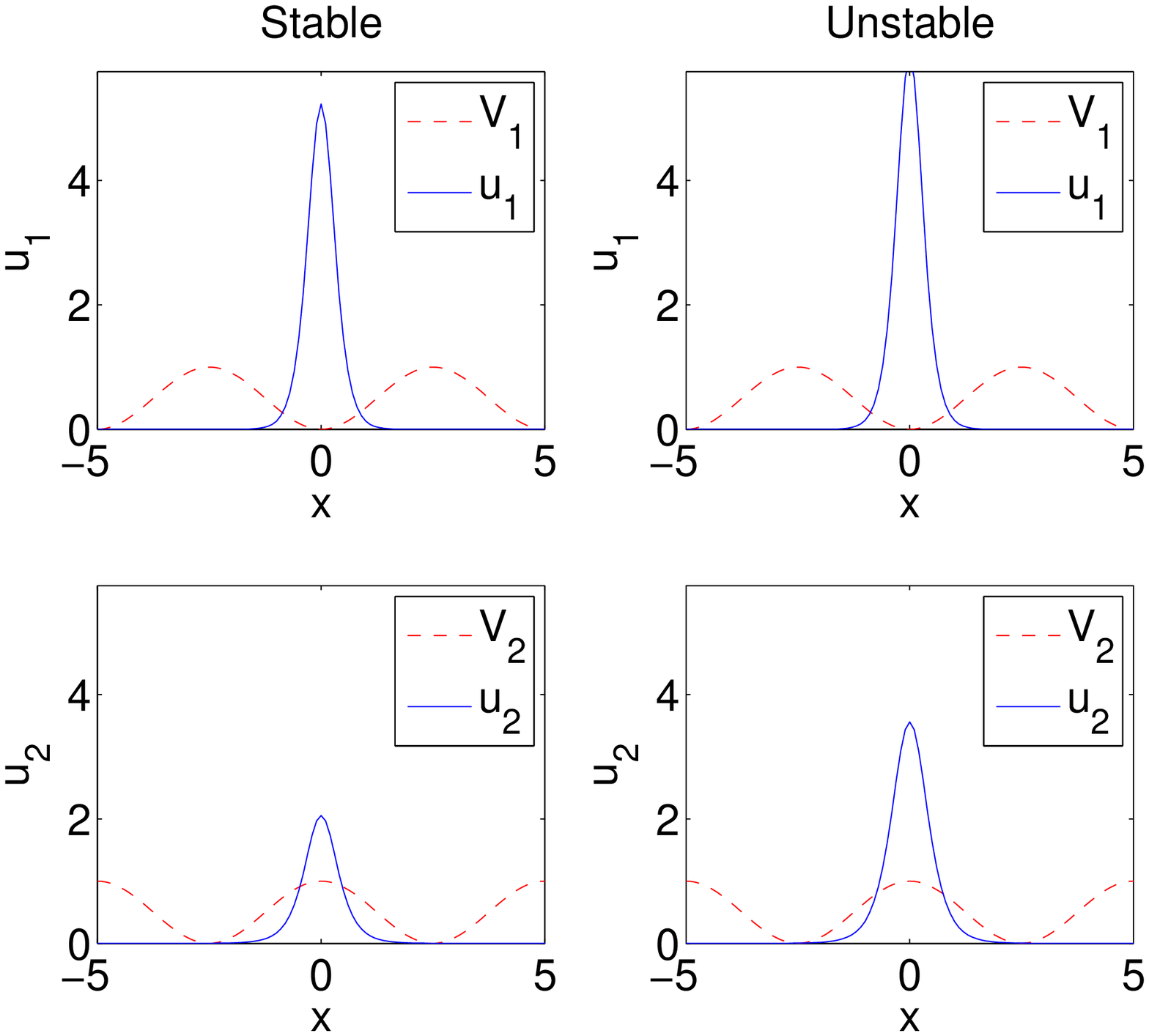}
\par
\includegraphics[width=7.3cm,height=6cm]{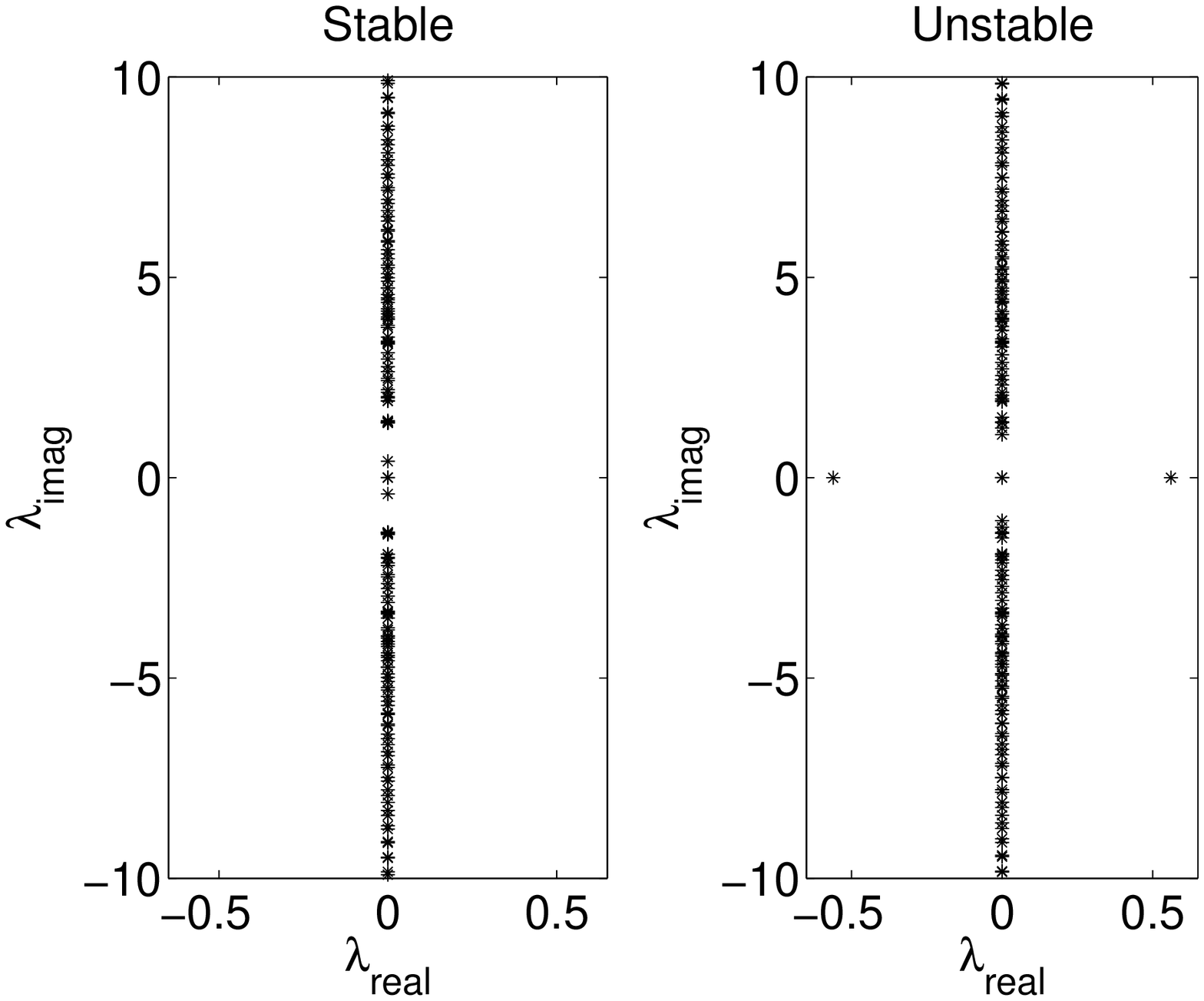}
\end{center}
\caption{(Color online) Profiles of both components of stable and unstable
solitons of the bright-bright type. The corresponding parameters in Eqs. (%
\protect\ref{eqn2}) and (\protect\ref{V}) are $V_{01}=V_{02}=1$, $\protect%
\mu _{1}=-3$, $\protect\mu _{2}=-1$, $g_{11}=-1$, $g_{22}=-1$, and $\left(
g_{12}\right) _{\mathrm{st}}=-0.4$ or $\left( g_{12}\right) _{\mathrm{unst}%
}=-0.04$, for the stable and unstable solitons, respectively. The solid and
dashed lines show the profiles of the components and of the periodic
potential, respectively. The bottom panel shows the respective spectral
planes of the stability eigenvalues, $\protect\lambda =\protect\lambda _{%
\mathrm{real}}+i\protect\lambda _{\mathrm{imag}}$, which identifies the left
and right solitons as stable and unstable ones, respectively.}
\label{fig1}
\end{figure}

\begin{figure}[tbp]
\includegraphics[width=7.3cm,height=6cm]{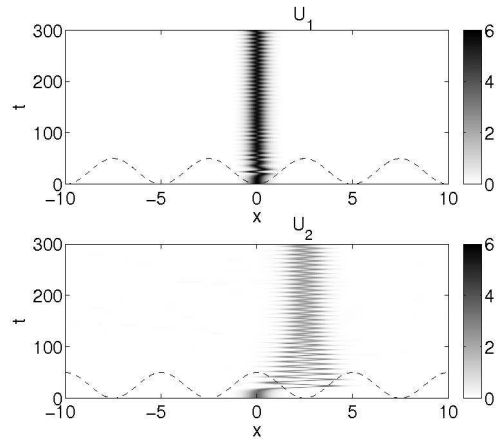}
\caption{(Color online) The dynamics of both components of an unstable
soliton complex of the bright-bright type is shown by means of density
space-time contour plots. The parameter setting is the same as in the
unstable case of Fig. \protect\ref{fig1}.}
\label{fig3}
\end{figure}

\begin{figure}[tbp]
\includegraphics[width=7cm,height=6cm]{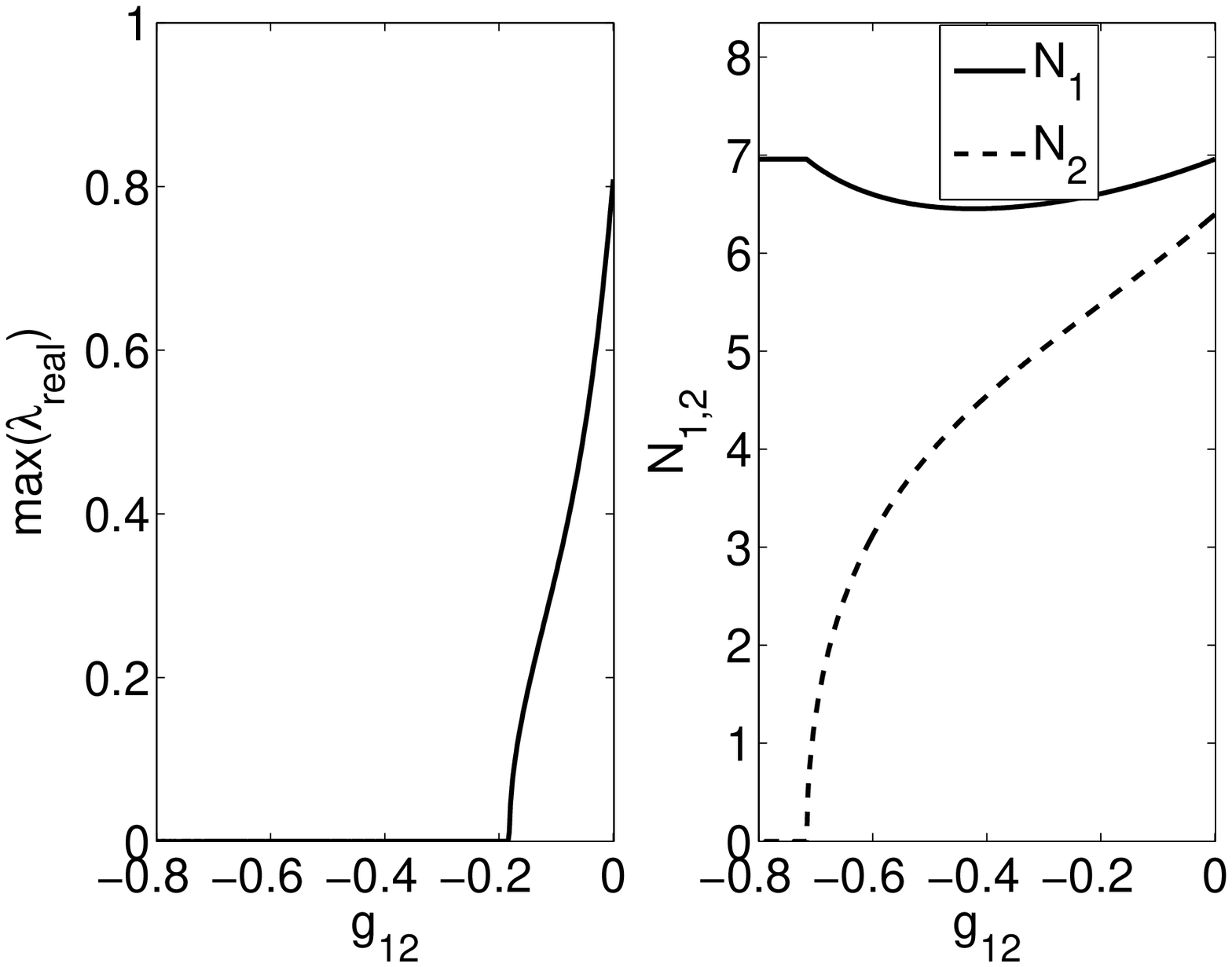}
\caption{(Color online) The change of the stability of the bright-bright
soliton (shown through the largest instability growth rate), and the
variation of the norms of both of it's components, as $g_{12}$ decreases
from $0$ to $-1$. Other parameters are $V_{01}=V_{02}=1$, $\protect\mu %
_{1}=-3$, $\protect\mu _{2}=-1$, $g_{11}=g_{22}=-1$.}
\label{fig4}
\end{figure}

\begin{figure}[tbp]
\caption{(Color online) The stability diagram for the soliton complexes of
the bright-bright type in the plane of $V_{01}=V_{02}\equiv V_{0}$ and $%
g_{12}$ (the stability is identified as per the maximum value of growth rate
$\protect\lambda _{\mathrm{real}}$ whose contour plot is shown). The other
parameters are $\protect\mu _{1}=-3$, $\protect\mu _{2}=-1$, $%
g_{11}=g_{22}=-1$. The black color means that the soliton complex is stable (%
$\protect\lambda _{\mathrm{real}}\equiv 0 $); the lighter the color, the
more unstable the soliton is. The curved line indicates the stability
border, as found from the numerical data.}
\label{fig5}\includegraphics[width=7cm,height=6cm]{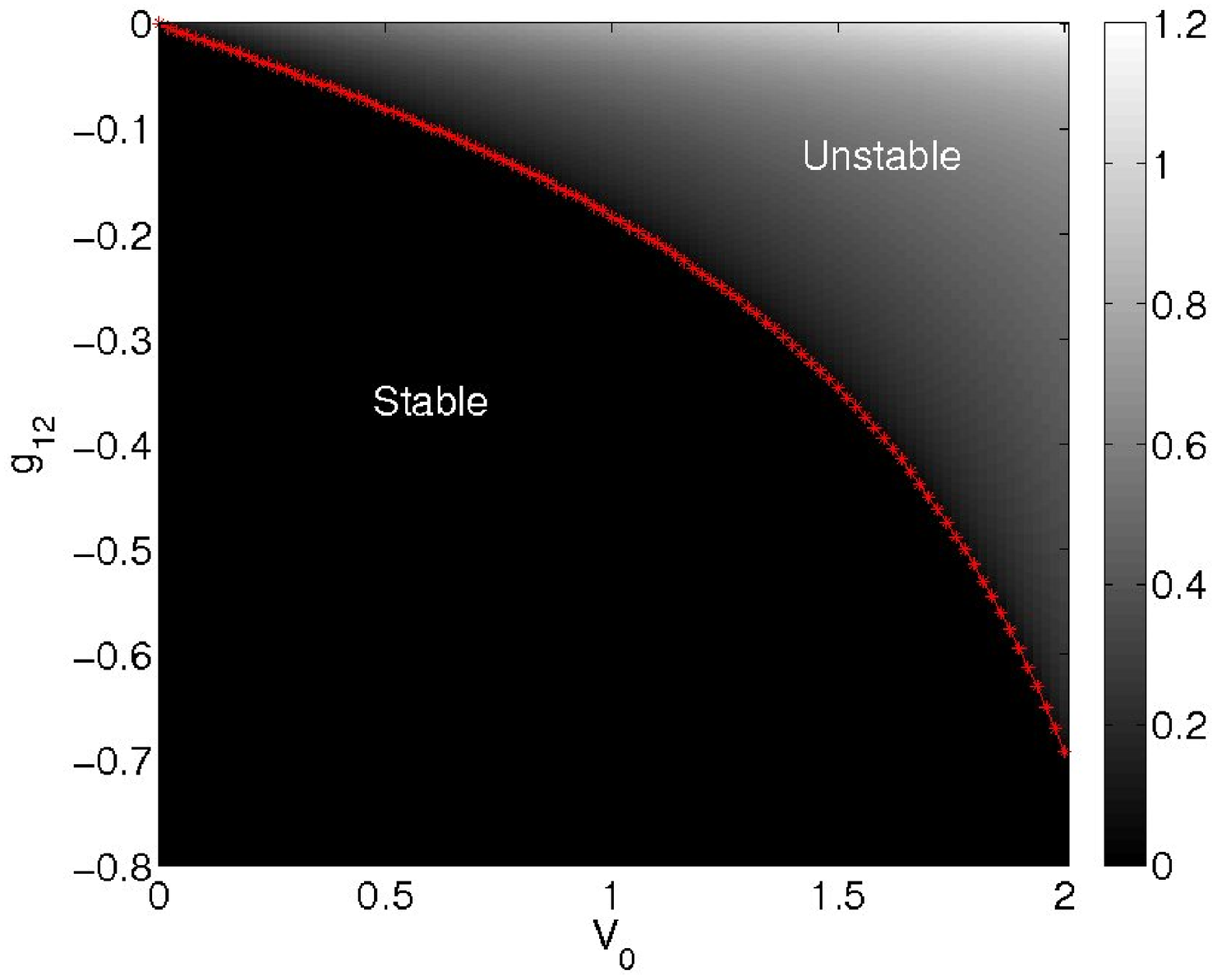}
% and predicted by the analytical
%approximation, see Eq. (\protect\ref{line}), respectively.}
\end{figure}

The OL potentials acting on the two species are assumed to be of the form
\begin{equation}
V_{1}(x)=V_{01}\sin ^{2}(kx),\ V_{2}(x)=V_{02}\cos ^{2}(kx),  \label{V}
\end{equation}%
which obviously corresponds to the phase shift of $\pi $ between the
sublattices. Figures \ref{fig1}-\ref{fig5} provide a systematic presentation
of the results, with a typical value of the wavenumber, $k=\pi /5$. First,
profiles of a pair of two-component solitons of the BB type, one unstable
(through a\ pair of real eigenvalues), and the other one stable, are
displayed in Fig. \ref{fig1}. In either case, both components of the soliton
are originally centered at $x=0$. In accordance with this, the component $%
U_{1}$, which is centered around its local potential minimum [see Eq. (\ref%
{V})] is a stable one, while $U_{2}$, whose original location coincides with
a local \textit{maximum} of the respective potential, $V_{2}(x)$, may be
unstable.

The evolution of the unstable soliton shown in Fig. \ref{fig3} (the one
pertaining to $g_{12}=-0.04$) was examined by means of direct simulations,
as shown in Fig. \ref{fig3}. It is seen that the stable component ($U_{1}$,
in this case), remains centered around the minimum of the corresponding
potential well, $x=0$, while the unstable component ($U_{2}$) is displaced
by the instability and ends up in an oscillatory state around a nearby
minimum of its respective potential, $x=\pi /\left( 2k\right) \equiv 5/2$.
%(the
%selection of $x=+5/2$, rather than $x=-5/2$, is a manifestation of the SSB).
This is a typical scenario of the instability development observed in the
present setting.

Results of a systematic analysis of the existence and stability of the
soliton family are presented in Figs. \ref{fig4} and \ref{fig5}. Figure \ref%
{fig4} shows a continuation of the family in $g_{12}$, the cross-coupling
interaction strength. At $g_{12}=0$, the configuration is obviously
unstable, due to the fact that one of the solitary waves ($U_{2}$) is
sitting on top of a potential maximum (see, e.g., Ref. \cite{kapitula} and
references therein for a rigorous investigation of the instability). As the
interaction between the two components gets enhanced, the stable first
component creates an effective attractive potential for the second
component, that eventually [in Fig. \ref{fig4}, this happens at $%
g_{12}<-0.19 $] overcomes the repulsive effect of the local potential $V_{2}$%
, and thus makes the configuration stable. Due to this mechanism, one may
speak about \textit{symbiotically stabilized} solitary waves. As the
continuation progresses towards more negative values of $g_{12}$, the second
component eventually disappears, and the resulting single-component bright
soliton is no longer affected by $g_{12}$.

Full results are summarized in the two-parameter soliton stability diagram
displayed of Fig. \ref{fig5}, with the grayscale indicating the magnitude of
the unstable eigenvalue. The stability border between the linearly stable
and linearly unstable regions is given by the curved (red in the online
version) line.

%We approximate this
%stability border (straight -blue in the online version- line in the figure)
%through the following considerations: we formulate an \textit{%
%effective potential} in the equation for $u_{2}$ in system (\ref{eqn2}), $V_{%
%{\rm eff}}(x)\equiv V_{02}\cos ^{2}(kx)+g_{12}|u_{1}|^{2}$. In the
%absence of the external potential, the single-component soliton in
%the first
%component is $u_{1}=\sqrt{2\left\vert \mu _{1}\right\vert }{\rm sech}%
%\left( \sqrt{2\left\vert \mu _{1}\right\vert }x\right) $. Inserting this in $%
%V_{{\rm eff}}(x)$,
%%we predict the blue line as a set of points at
%%which the local maximum of $V_{{\rm eff}}(x)$ switches into a
%we predict the change of dynamical behavior as occurring when
%the sign of the potential at the center of the soliton changes, namely when
%%minimum, i.e.,
%%the equilibrium position becomes stable, for given $\mu _{1}$:%
%\begin{equation}
%\left\vert g_{12}\right\vert =V_{02}/\left( 2\mu _{1}\right) .
%\label{line}
%\end{equation}%
%Naturally, the accuracy of this prediction should increase with the decrease
%of $\left\vert g_{12}\right\vert $, i.e., decrease of $V_{0}$, which is
%indeed observed in Fig. \ref{fig5}

\subsection{Solitons of the gap-bright type}

\begin{figure}[tbp]
\begin{center}
\includegraphics[width=7.3cm,height=6cm]{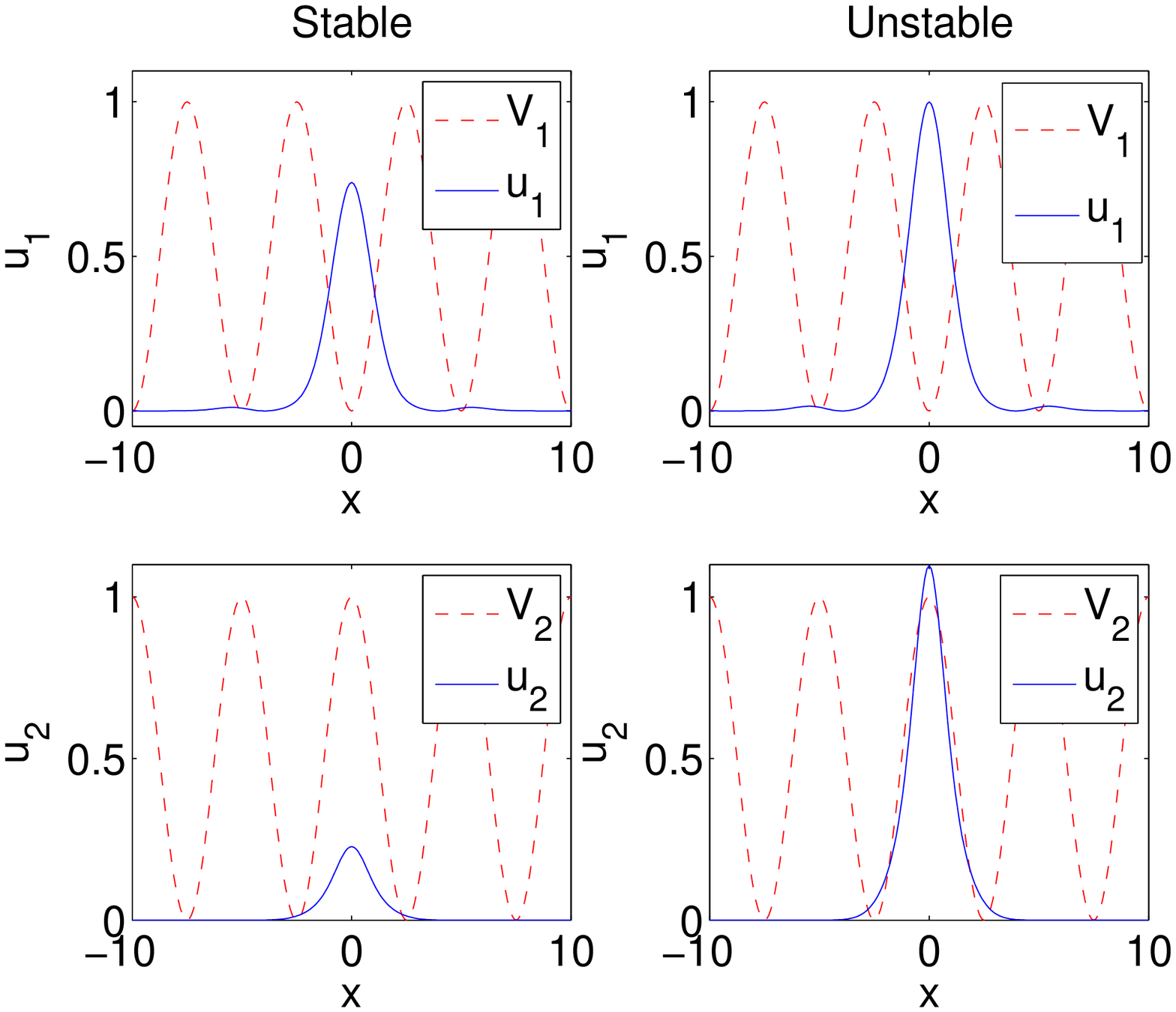}
\par
\includegraphics[width=7.3cm,height=6cm]{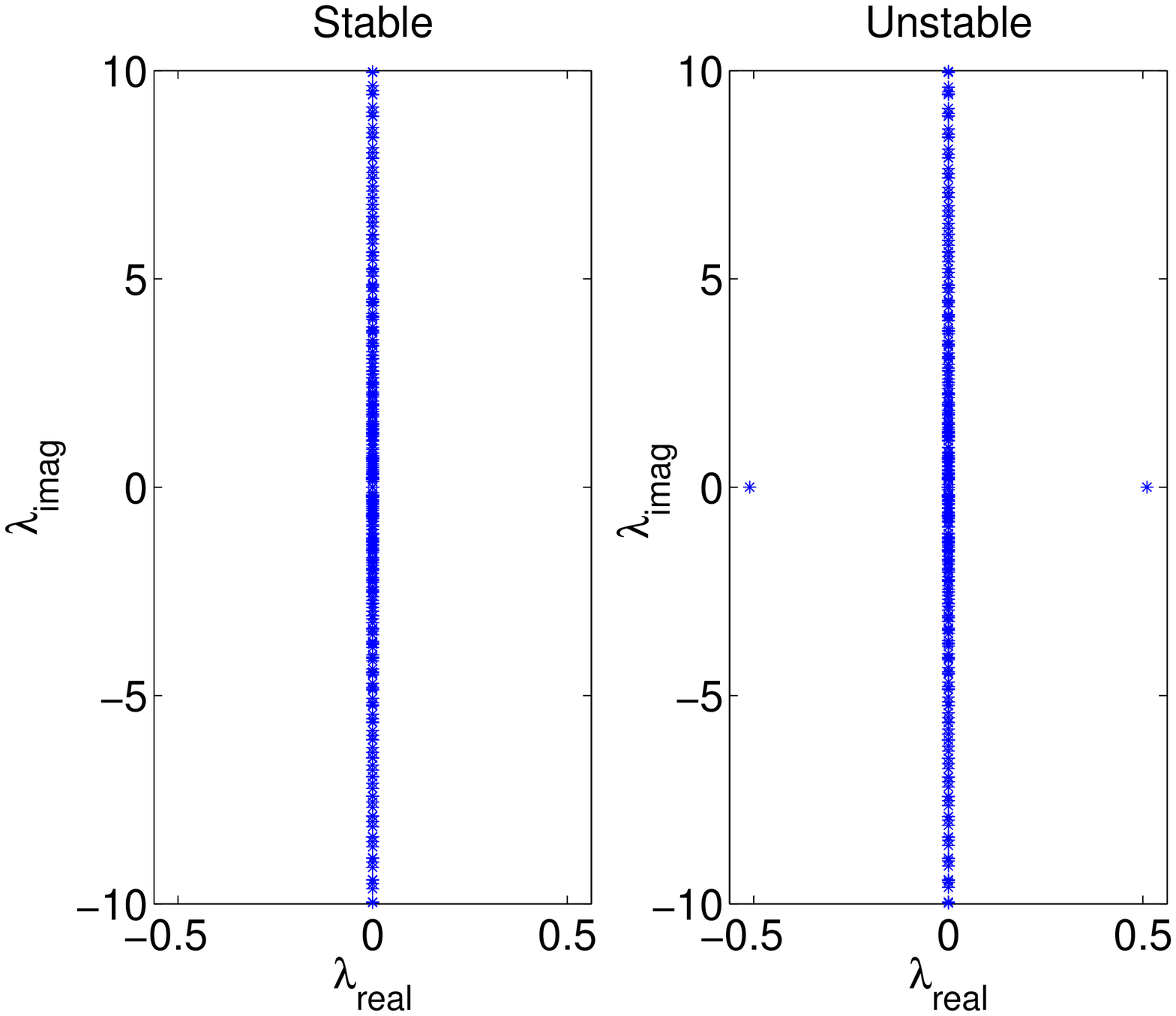}
\end{center}
\par
%Parameter settings
%are: $v_1=0$, $v_2=v_0cos(kx)^2$,$v_0=1$, $\lambda1
%=-3$,$\lambda2=-1$,$g_{11}=-1$,$g_{22}=-1$, $g_{12}=-0.04$ when
%system is unstable and $g_{12}=-0.4$ when system is stable. Solid
%line shows the profile of both components and Dashed line shows the
%periodic potential
\caption{(Color online) Profiles of both components of gap-bright soliton
complexes in stable and unstable states. The parameters are: $%
V_{01}=V_{02}=1 $, $\protect\mu _{1}=0.6$, $\protect\mu _{2}=-0.3$, $g_{11}=1
$, $g_{22}=-1$, and $\left( g_{12}\right) _{\mathrm{st}}=-2.0$ , $\left(
g_{12}\right) _{\mathrm{unst}}=-0.675$, for the stable and unstable
solitons, respectively.}
\label{fig1a}
\end{figure}

%We now proceed to show analogous results in the case of $V_1(x)=0$,
%i.e., in the absence of a periodic potential in the first component.
%The corresponding figures for this case are \ref{fig1a}-\ref{fig5a},
%which are done for similar parametric regimes as the ones above
%in the presence of the optical lattice potential in the first
%component. The key observation to make here (already shown
%in Fig \ref{fig1a}, but clearly evident in Fig. \ref{fig4a}) is
%that there is {\it no symbiotic stabilization} in this case.
%The system becomes stabilized {\it only} when the second
%component completely disappears.
Typical examples of stable and unstable stationary states of the GB type are
shown in Fig. \ref{fig1a}. In this setting, the potentials are again taken
as in Eq. (\ref{V}). The stable component, $U_{1}$, is always of the gap
type (i.e., with the repulsive intrinsic interaction, $g_{11}>0$), while the
the self-attractive (``bright") component, $U_{2}$, with $g_{22}<0$, may be
unstable.

\begin{figure}[tbp]
\begin{center}
\includegraphics[width=7.3cm,height=6cm]{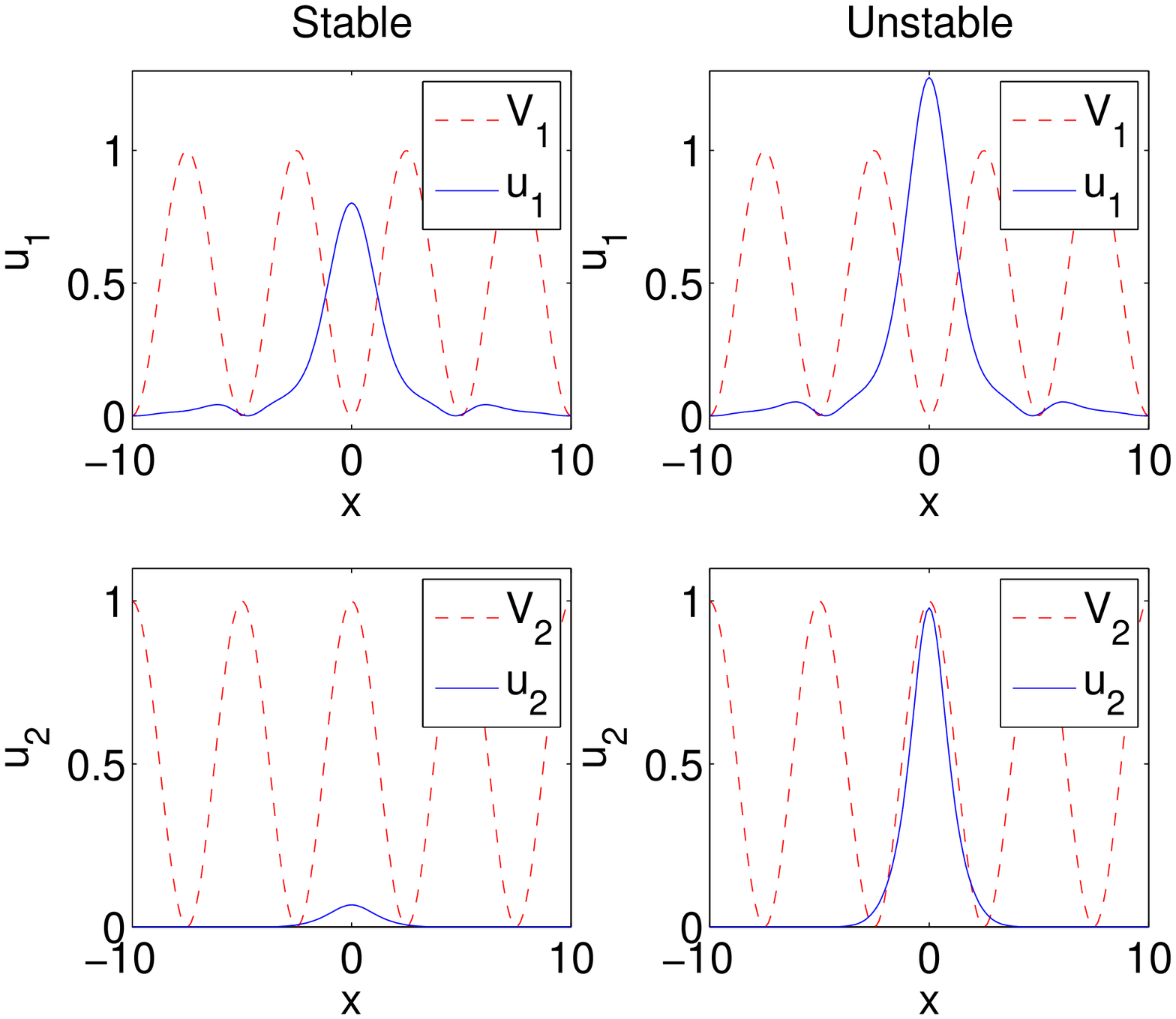}
\end{center}
\caption{(Color online) The same as in Fig. \protect\ref{fig1a}, except for $%
\protect\mu _{1}=0.88,$ $\protect\mu _{2}=-0.4,$ and $\left( g_{12}\right) _{%
\mathrm{st}}=-2.00$, $\left( g_{12}\right) _{\mathrm{unst}}=-0.675.$}
\label{fig1b}
\end{figure}

The examples displayed in Fig. \ref{fig1a} correspond to $\mu _{1}=0.6$,
which actually falls in the middle of the first bandgap in the OL-induced
linear spectrum of the first component. For this reason, the gap-type
component ($u_{1}$) does not feature conspicuous tails \cite{gap-symbio,SKA}%
. Taking a larger value of the chemical potential, for instance, $\mu
_{2}=0.88$, which is chosen closer to the edge of the bandgap, we can
generate the gap-components of the soliton with more salient tails, see Fig. %
\ref{fig1b}.

\begin{figure}[tbp]
\begin{center}
\includegraphics[width=7.3cm,height=6cm]{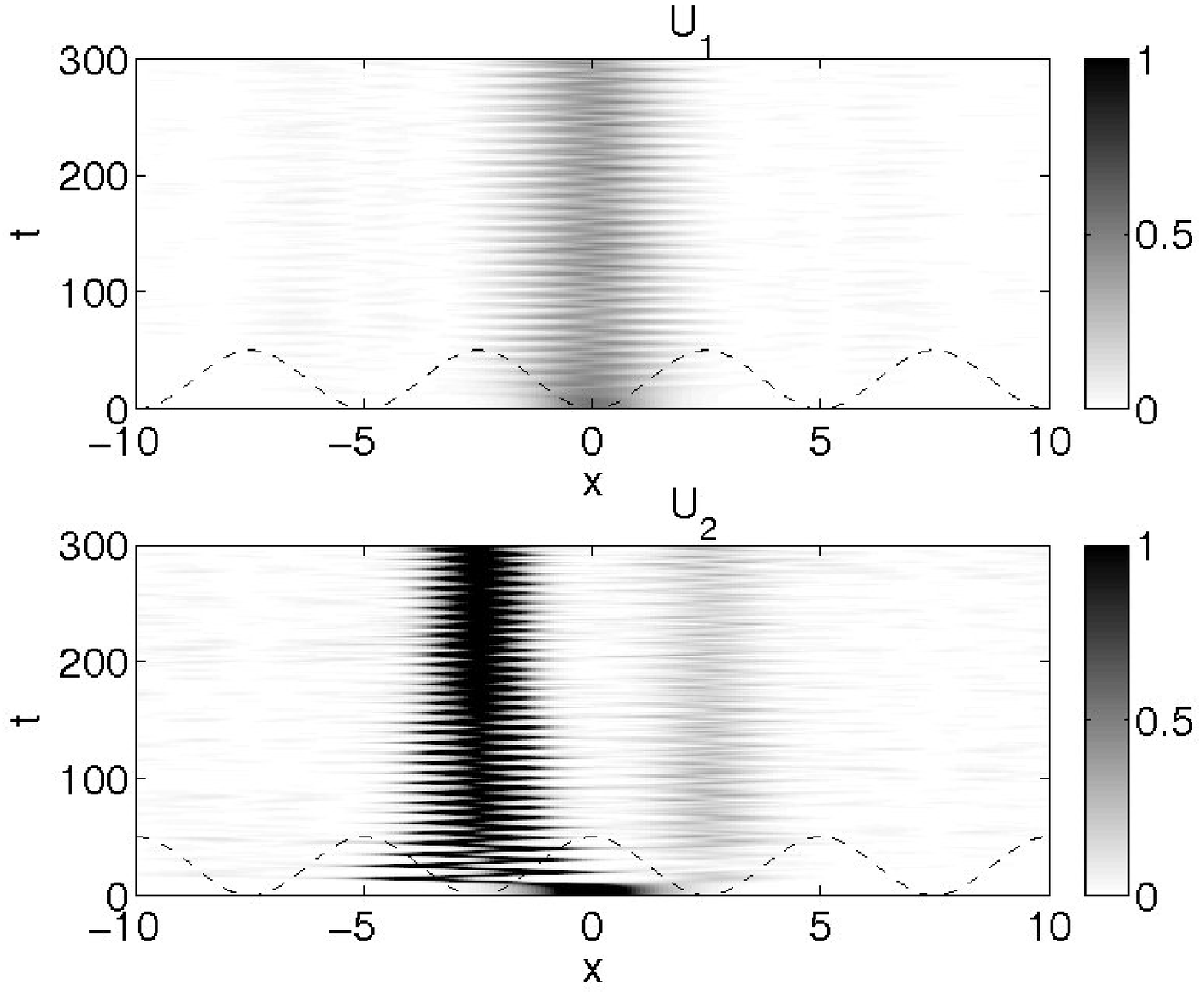}
\par
\includegraphics[width=7.3cm,height=6cm]{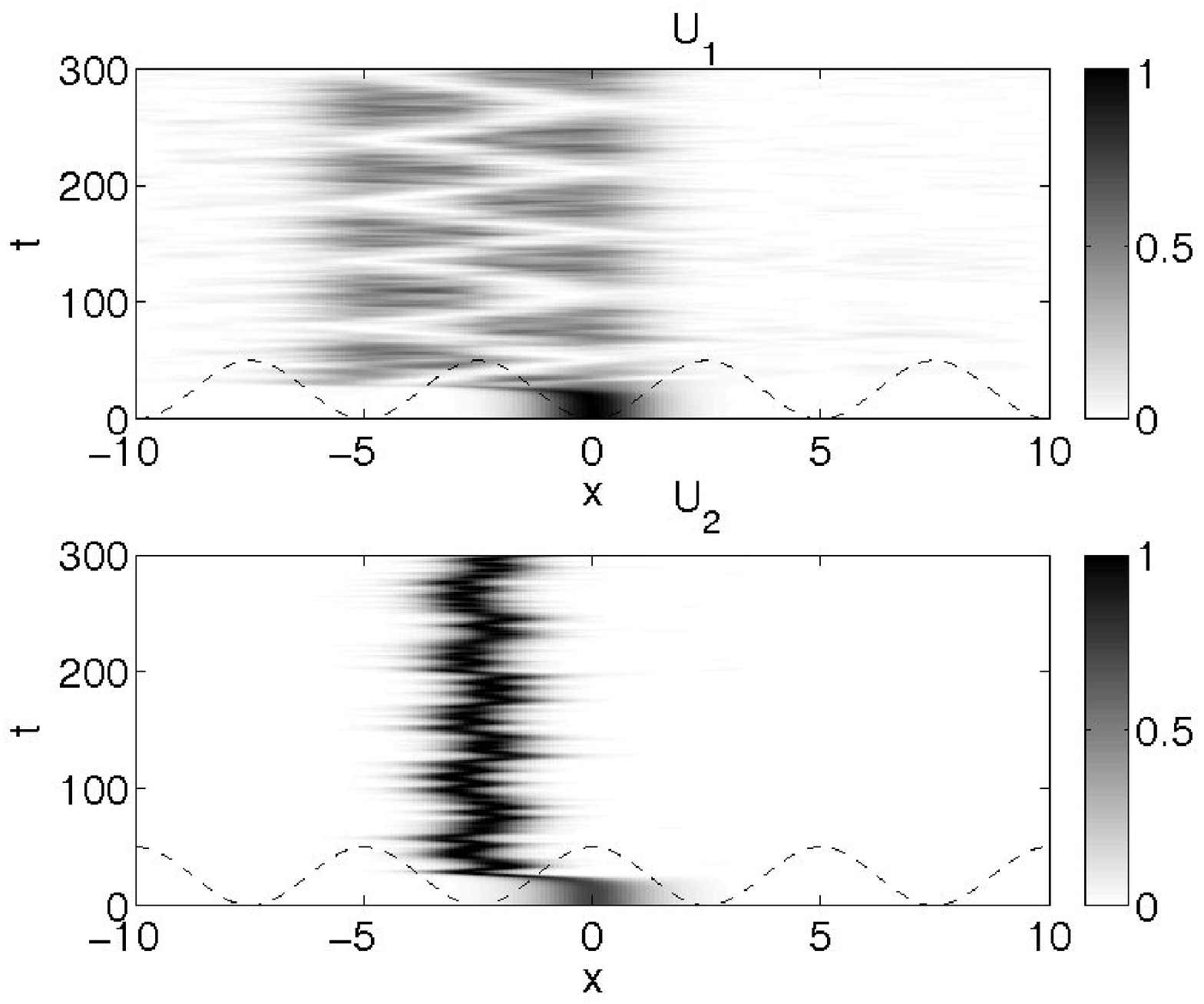}
\par
\includegraphics[width=7.3cm,height=6cm]{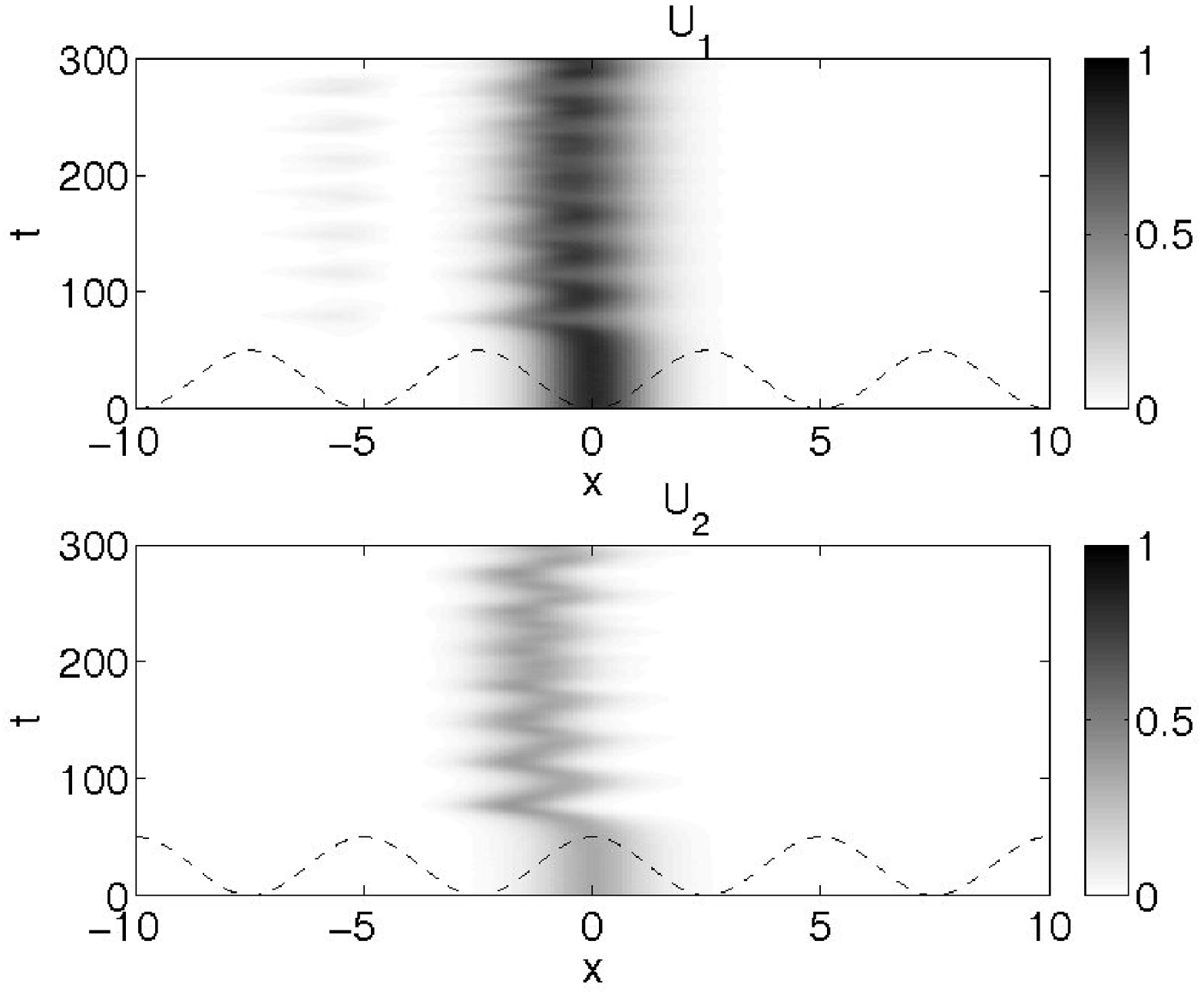}
\end{center}
\caption{(Color online) The evolution of an unstable soliton complex of the
gap-bright type. Parameters are the same as in the former pictures
displaying the gap-bright complexes, except that $g_{12}=-0.15$, $%
g_{12}=-1.00$, and $g_{12}=-1.65$, in the top two rows, third and fourth
rows, and bottom two rows, respectively.}
\label{fig3a}
\end{figure}

Simulations of the evolution of unstable complexes of the GB type, as shown
in Fig. \ref{fig3a}, reveal an essential difference from the
instability-development pattern that was observed in the previous case (for
the two-component solitons of the BB type). In that case, independently of
the specific value of $g_{12}$, the stable component (which was also $U_{1}$%
) remained centered around $x=0$, while the unstable one, $U_{2}$,
oscillated around either of the two nearby minima of its respective
potential, see Fig. \ref{fig3}. On the other hand, component $U_{2}$ in the
case of unstable GB solitons \emph{splits} into two \emph{unequal} parts, if
the interaction between the two components is weak (see, e.g., an example
for $g_{12}=-0.15$ in the top two rows of Fig. \ref{fig3a}). The two
splinters end up oscillating around different nearby potential minima, $%
x=\pm 5/2$. %(the difference in their norms is another manifestation of the
%instability-induced SSB).
As $|g_{12}|$ rises to intermediate values, such as $g_{12}=-1$, the
unstable component $U_{2}$ of the GB soliton ceases to feature splitting,
but rather oscillates around one of the minima, while the first (stable)
component, $U_{1}$, is found to oscillate between its original position, $%
x=0 $, and an adjacent potential minimum, $x=-2\pi /(2k)\equiv -5$ (in fact,
$U_{1}$ oscillates around the average position of component $U_{2}$). These
scenarios of the instability development are displayed in the third and
fourth rows of Fig. \ref{fig3a}). As the interaction between the components
becomes still stronger -- for instance, at $g_{12}=-1.65$ -- both components
feature approximately \textit{synchronous} oscillations (i.e., the soliton
complex keeps its integrity, due to the strong attraction between its
constituents) in the range of $-2.5<x<0,$ that is, between the minimum of
potential $V_{1}(x)$/maximum of $V_{2}(x)$ and the adjacent maximum of $%
V_{1}(x)$/minimum of $V_{2}(x)$.

%\begin{figure}[tbp]
%\begin{center}
%\includegraphics[width=7.3cm,height=6cm]{eigen_withoutv1.eps}
%\end{center}
%\caption{(Color online) Eigenvalue spectrum of unstable system and
%stable system when $v_1$ is zero; Parameter settings are the same as
%in figure1 } \label{fig2a}
%\end{figure}

%The position of
%$x=-5$ is a maximum potential point for the first component and the
%minimum potential point for the second component. The specific
%phenomenon happens because of the humps of the gap soliton.

\begin{figure}[tbp]
\includegraphics[width=10cm,height=5cm]{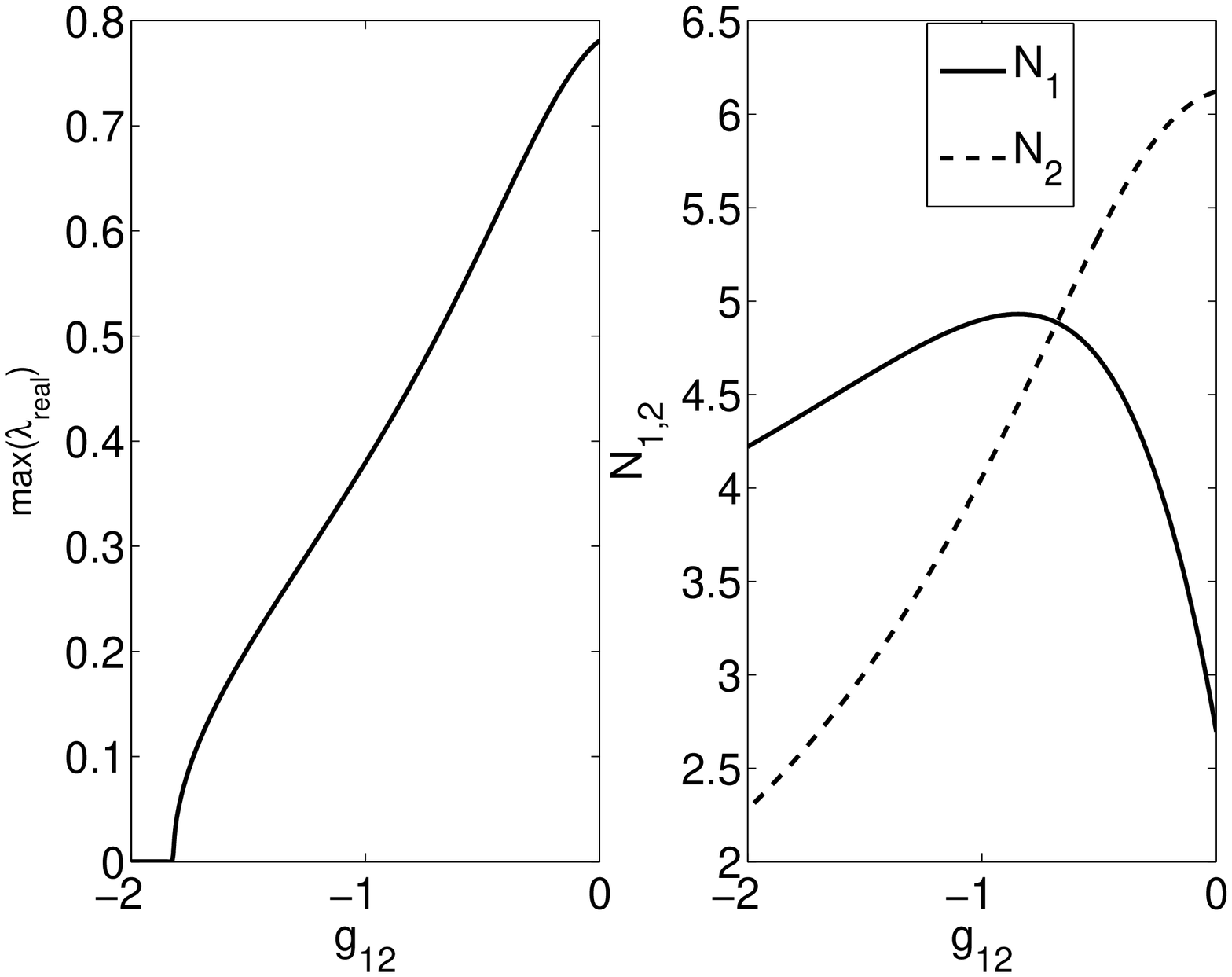}
\caption{(Color online) Left panel: the largest instability growth rate for
solitons of the gap-bright type. Right panel: the evolution of the norms of
both components with the decrease of $g_{12}$ from $0$ to $-2$. The
parameters chosen are $V_{01}=V_{02}=1$, $\protect\mu _{1}=0.6$, $\protect%
\mu _{2}=-0.3$, $g_{11}=1$, $g_{22}=-1$. }
\label{fig4a}
\end{figure}

Performing the continuation in the interaction-strength parameter, $g_{12}$,
we have generated characteristics of the GB-soliton family, a typical
example of which is displayed in Fig. \ref{fig4a}, cf. Fig. \ref{fig4} for
complexes of the BB type. Further, collecting the results for different
values of $V_{02}$ [while $V_{01}$ is fixed, see Eq. (\ref{V})], we have
produced the stability diagram for the GB solitons, as shown in Fig. \ref%
{fig5a}, cf. its counterpart for the solitons of the BB type in Fig. \ref%
{fig5}. It is worthy to note that the critical value of $g_{12}$ at the
stability border decreases almost linearly as $V_{02}$ increases.
%the stabilization of the two
%component system (even as $V_0 \rightarrow 0$) does {\it not} occur
%for $g_{12} \rightarrow 0$, but for a finite value of $g_{12}$. This
%stabilization in fact has always been observed in our numerical
%results to occur when $g_{12}$ is strong enough so that the second
%branch $u_2(x) \equiv 0$. I.e., as indicated above in the absence of
%the first-component potential, the symbiotic stabilization effect
%discussed previously is also absent.

\begin{figure}[tbp]
\includegraphics[width=7cm,height=6cm]{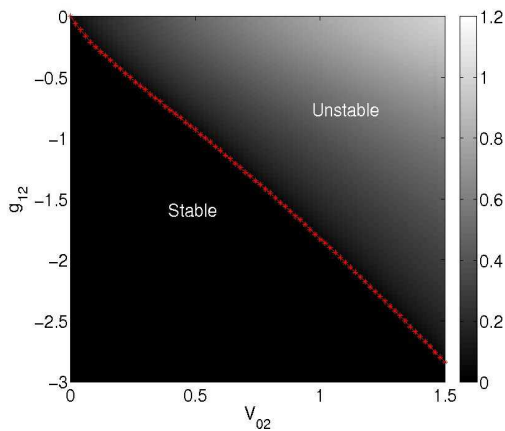}
%parameter settings are : $v_1=0$,
%$v_2=v_0cos(kx)^2$, $v_0=1$, $\lambda1 =-3$, $\lambda2=-1$,
%$g_{11}=-1$, $g_{22}=-1$. Dark color means that system is stable at
%fixed $v_0$ and $g_{12}$, the lighter of the color, the more unstable
%the system is.  Red curve shows the simulated relation of critical
%$g_{12}$ and $v_0$.
\caption{(Color online) The stability diagram for solitons of the gap-bright
type in the plane of $V_{02}$ and $g_{12}$. The notation is the same as in
Fig. \protect\ref{fig5}. Other parameters are fixed: $V_{01}=1$, $\protect%
\mu _{1}=0.6$,$\ \protect\mu _{2}=-0.3$, $g_{11}=1$, $g_{22}=-1$.}
\label{fig5a}
\end{figure}

\section{Conclusion}

The objective of this work was to extend the model of binary BEC mixtures to
the case when the two components are trapped by mismatched OLs. The analysis
was focused on the case of the strongest mismatch, with the sublattices of
opposite signs. The intrinsic nonlinearity in one species was
self-attractive, while in the other one it might have either sign; however,
the inter-species interaction was always attractive. Due to the opposite
signs of the two effective OL potentials, the bound state of two solitons
could be unstable, as the second component of this state had to be centered
around a local potential maximum. In both cases when the nonlinearity in the
stable component (the one centered around the respective potential minimum)
is self-attractive or self-repulsive (while the other species is
self-attractive), we have identified stability regions for the soliton
complexes, varying the strength of the inter-species attraction and of the
OL depth. %In the former
%(``bright-bright", BB) case, we have also developed a simple
%analytical approximation for the stability border.
In cases when the soliton complex is unstable, direct simulations have made
it possible to identify basic scenarios of the instability development,
which amount to the spontaneous symmetry breaking (SSB) and resulting
oscillations of one or both components. In the case of the soliton complex
of the ``gap-bright" (GB) type, the unstable (``bright") component may also
potentially split into two unequal parts, which oscillate around different
local potential minima. It is relevant to mention that it was not possible
to implement a different type of the instability, where the gap-type
component (the intrinsically self-repulsive one) would be originally set in
an unstable position with respect to its sublattice, as a complex of such a
type could \emph{never} be constructed (even as a formal solution). For the
same reason, we did not consider complexes of the gap-gap type.

As concerns possible extensions of this work, an interesting issue is to
consider the stability and dynamics of two-component solitons in two
dimensions, that would be supported by two square lattices with the phase
shift of $\pi $ in both directions $x$ and $y$. In that case, it would be
interesting to examine whether the unstable component may perform
two-dimensional quasi-circular motion around the pinned stable one.

\end{document}